\documentclass{iaus}
\usepackage{graphicx}

\title[Multiple populations in GCs] 
{The quite complex ``Simple Stellar Populations" of Globular Clusters}

\author[A. Bragaglia]   
{Angela Bragaglia$^1$
}

\affiliation{$^1$INAF-Osservatorio Astronomico di Bologna,\\
via Ranzani 1, 40127 Bologna, Italy \\
email: {\tt angela.bragaglia@oabo.inaf.it}}

\pubyear{2010}
\volume{268}  
\jname{Light elements in the Universe}
\editors{C. Charbonnel, M. Tosi, F. Primas \& C. Chiappini, eds.}
\begin{document}

\maketitle

\begin{abstract}
There is compelling observational evidence that globular clusters (GCs) are
quite complex objects. A  growing body of photometric results indicate that the
evolutionary sequences are not simply isochrones in the observational plane -as
believed until a few years ago-  from the main sequence, to the subgiant, giant,
and horizontal branches.  The strongest indication of complexity comes however
from the chemistry, from internal  dispersion in iron abundance in a few cases,
and in light elements (C, N, O, Na, Mg, Al, etc.) in {\em all}  GCs.  This
universality  means that the complexity is {\em intrinsic} to the GCs and is
most probably related to their formation mechanisms.  The extent of the
variations in light elements abundances  is  dependent on the GC mass, but mass
is not the only modulating factor; metallicity, age, and possibly orbit can play
a role. Finally, one of the many consequences  of this new way of looking at GCs
is that their stars may show different He contents.
\end{abstract}

\firstsection 
\section{It's not so simple}

Globular Clusters (GCs) have long been considered the best approximations of a
{\em Simple Stellar Population} (\cite[Renzini \& Buzzoni 1986]{ssp}), and this
may still be valid for some purposes. However, they are not truly simple. We
know that GCs contain a fraction of binaries (e.g., \cite[Meylan \& Heggie
1987]{mh87}). But now we also know that their stars are not strictly coeval; for
old clusters, like the Galactic globulars, the age differences are so small to
be hardly detectable, but the same is not true for the Magellanic Clouds ones.
And we also know that the initial chemical composition of the stars we presently
observe was not the same.

There is a growing body of observational evidence of the complexity of GCs, both
from photometry (with multiple, or split, or wide sequences) and from
spectroscopy (with large differences in light elements and even, in a few cases,
with spreads in metallicity).

Of course, $\omega$ Cen is the first example that comes to mind, even if for 
its characteristics (or better, because of its characteristics)  it has often
been labelled as the nucleus of an ancient dwarf spheroidal galaxy. However,
$\omega$ Cen is only the tip of the iceberg and  there are many other
interesting cases, like M54, which lies in the nucleus of the Sagittarius  dwarf
galaxy (\cite[Ibata et al. 1994]{ibata}, \cite[Bellazzini et al.
2008a]{bell08a}), and which resembles $\omega$ Cen in several aspects. But also
more ``normal", lower mass clusters show peculiarities, like NGC~2808 which,
among many other oddities, presents three well separated main sequences
(\cite[Bedin et al. 2004]{bedin04}, \cite[Sollima et al. 2007]{sollima07}). I
will come back to these three objects later.

\begin{figure}[]
\begin{center}
\includegraphics[bb=20 140 580 710, clip,width=6cm]{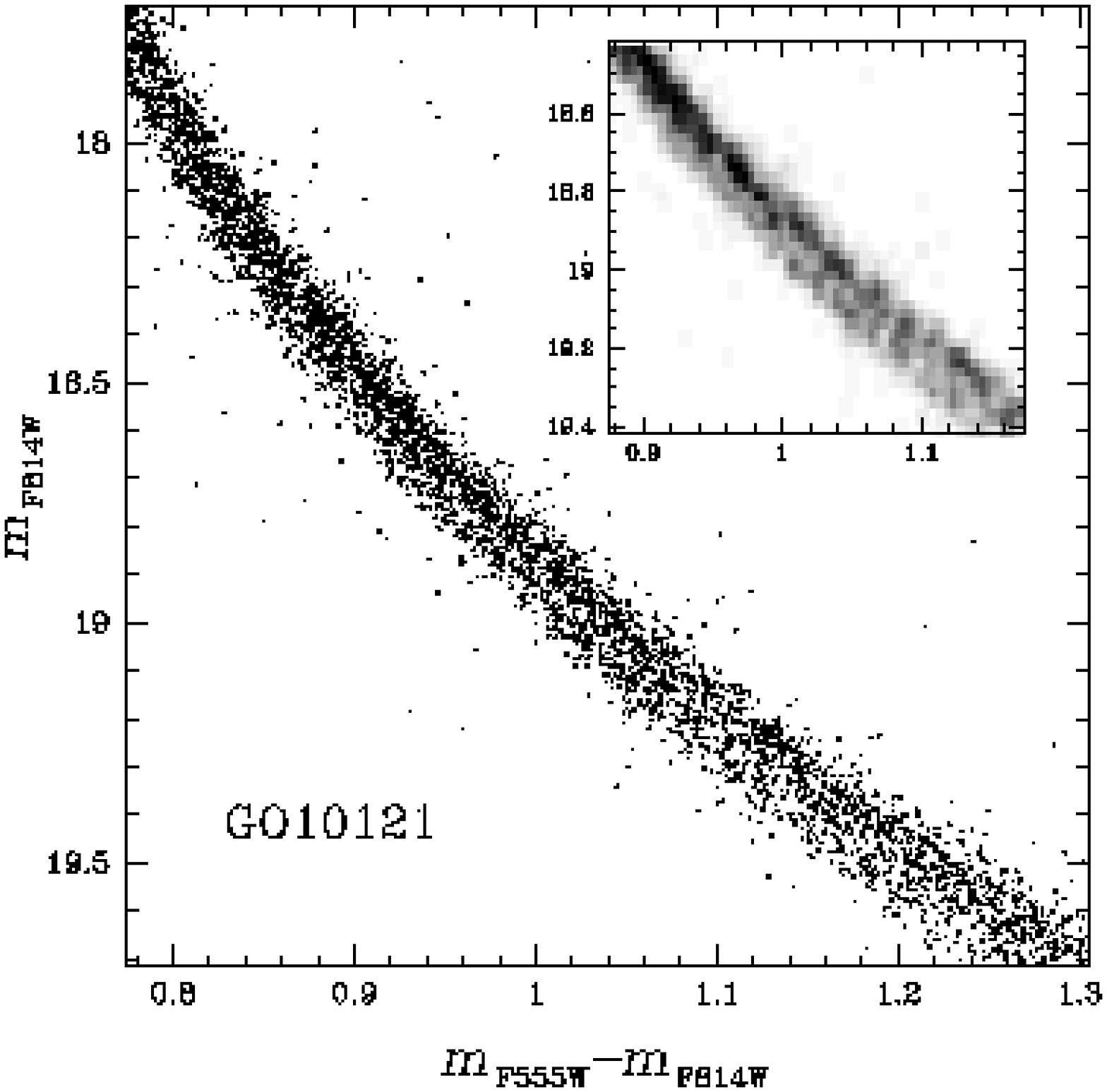} 
\includegraphics[bb=110 240 450 610, clip,width=6cm]{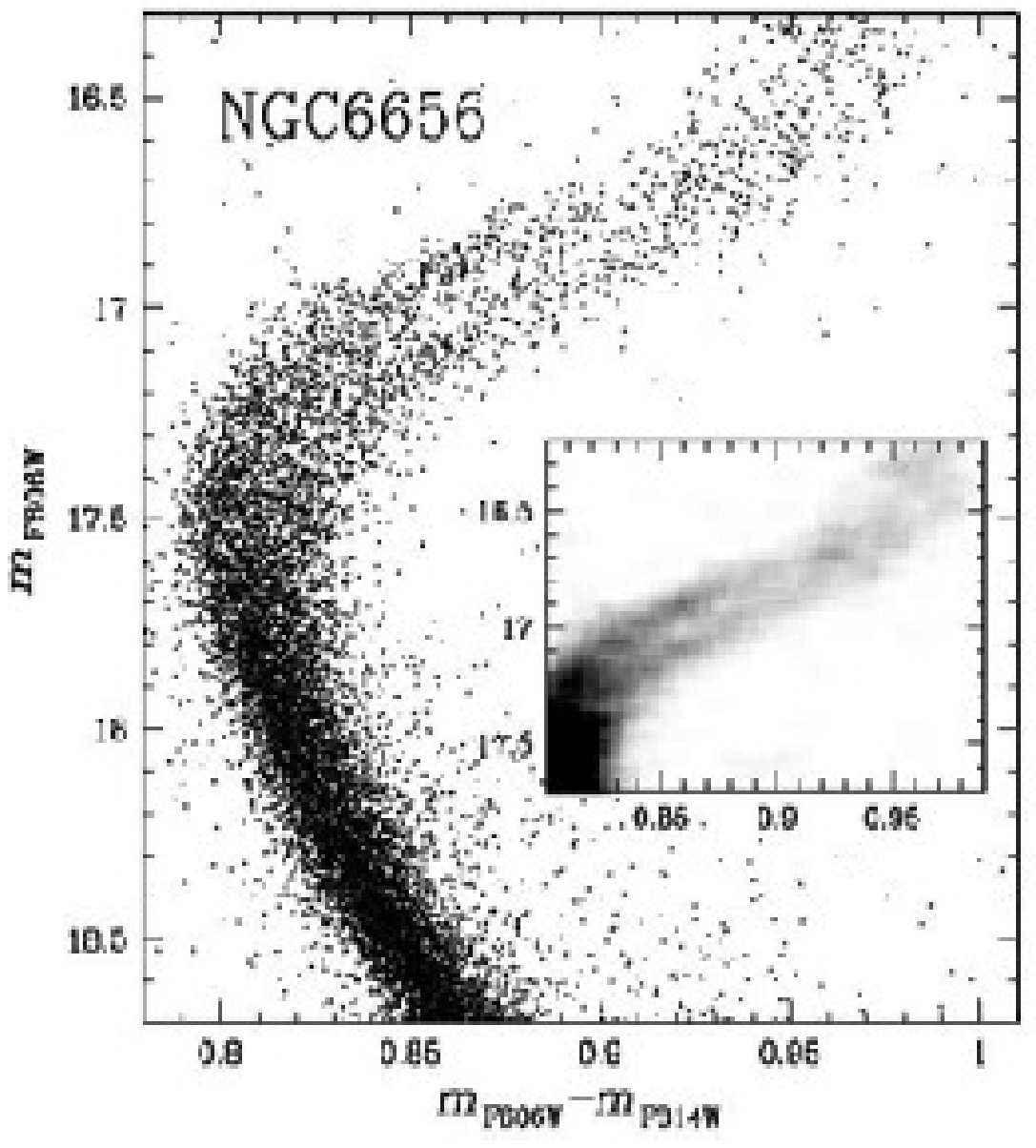} 
\caption{Examples of split main sequences or subgiant branches recently found
in Galactic GCs using the high precision photometry of ACS@HST: NGC 6752
(\cite[Milone et al. 2009]{milone6752}), and M22 (courtesy of A. Milone).}
\label{figsplit}
\end{center}
\end{figure}

Among other evident  examples we find for instance M22 (NGC 6656)  which had
long been suspected to have  a dispersion in metallicity (from photometry, e.g. 
\cite[Hesser et al. 1977]{hesser} or spectroscopy, e.g. \cite[Brown \&
Wallerstein 1992]{bw92}, but see also \cite[Ivans et al. 2004]{ivans}) and that
also displays a split in the SGB and RGB (see Fig. \,\ref{figsplit}), or
NGC~1851 with its split SGB (\cite[Milone et al. 2005]{milone05}) and RGB
(\cite[Han et al. 2009]{han}), and its anomalous chemical abundances (e.g.
\cite[Yong et al. 2009]{yong}).

Interestingly, thanks to the exquisite precision of the ACS camera on HST (whose
photometry can reach the depth and precision  to detect even small colour and
magnitude differences), more clusters were found to show wide or even split
evolutionary sequences, from the main sequence and up to the SGB and RGB.  The
newest entries are 47 Tuc, where \cite[Anderson et al. (2009)]{anderson} find
both a split SGB and a wide faint main sequence, and  NGC 6752, unsuspected
until now, which shows a split main sequence (\cite[Milone et a.
2009]{milone6752}); for both, see Fig. \,\ref{figsplit}. It seems that more and
more GCs are beginning to unveil their complex photometric sequences; for  a
more extended discussion, see the earlier reviews by \cite[Piotto (2008,
2009)]{p808}\cite[]{p809}. 
 
Even if the evidence from photometry is  the easiest to see, even for
non-specialists, the strongest proof that GCs harbour at least two stellar
generations comes from  spectroscopy.  Not only we may see stars with the same
evolutionary status but with very different chemical composition in GCs, at
least for some light elements, but also this situation is not limited to a few
``freaks", as $\omega$ Cen or M22\footnote{Recently, 
the claim of a dispersion in metallicity in M22 has gained
substance, and two papers appeared. One is based on a sample of 17 stars
studied with high resolution spectra (\cite[Marino et al. 2009]{marino09}), and indicates differences both in metallicity and
heavy elements. The second paper, by \cite[Da Costa et al. (2009)]{dacostam22}, presents a larger sample, but metallicity
is derived from the Calcium triplet. Both show a bimodality in metallicity.}
were considered for a long time.  The chemical signatures we used to call
``anomalies" are widespread and show up in {\em all} clusters  studied.
Bimodality -and anticorrelation- in CN and CH strengths, or anticorrelations
between other light elements, like Na and O, or Mg and Al, have long been
observed in evolved stars, where they could be explained, although with some
difficulties, considering extra-mixing episodes (see e.g. the reviews by
\cite[Smith 1987]{smith} and \cite[Kraft 1994]{kraft} where the extra-mixing
$vs$ the primordial-enrichment hypotheses are discussed).

However, these same so-called chemical anomalies were later found also in non
evolved, main sequence stars (e.g., \cite[Cannon et al. 1998]{cannon} found a
bimodality in CN, CH in 47 Tuc, and other authors in many other GCs, see the
review by \cite[Gratton et al. 2004]{araa}). 

Once established that the Na-O anticorrelation was present also in unevolved
stars (\cite[Gratton et al. 2001]{gra01} found it for the first time in NGC 6752
and NGC 6397, followed by \cite[Ramirez \& Cohen 2002]{rc02} in M71, and by
\cite[Carretta et al. 2004]{carretta04} in 47 Tuc),  it became clear that
another explanation was required.   These chemical signatures are  the result of
H-burning at high temperature (ON, NeNa, MgAl cycles: see \cite[Denisenkov \&
Denisenkova 1989]{dd}, \cite[Langer et al. 1993]{langer}); the resulting
chemical patterns cannot be produced in low mass, main sequence stars like those
we are presently seeing in GCs. So the typical  chemical signature of high N, 
Na, Al, and low C, O, Mg {\em  must have originated in a previous generation of
more massive stars, that polluted the gas from which part of the GC stars formed
later}.

Perhaps the most famous pair of elements showing such variations,
anti-correlated with each other, are O and Na. This Na-O anticorrelation was
first found among cluster giants,  mainly by the Lick-Texas group, by Kraft,
Sneden, and many collaborators. They studied two-three tens of targets per
cluster, observing stars one by one; for a review of their work, see \cite[Kraft
(1994)]{kraft} and \cite[Gratton et al.  (2004)]{araa}. The availability of
efficient spectrographs at 8-10m telescopes, and their multi-object capabilities
have permitted to extend this kind of study both to faint, unevolved stars and
to much larger samples.  About three years ago, there were about 200 giants
studied in literature, and about 50 unevolved stars, as shown in \cite[Carretta
et al. (2006)]{carretta06} in their presentation of the Na-O anticorrelation in
about 100 stars in NGC 2808, studied with FLAMES@VLT. 

\begin{figure}[]
\begin{center}
\includegraphics[bb=10 220 560 660, clip,width=13cm]{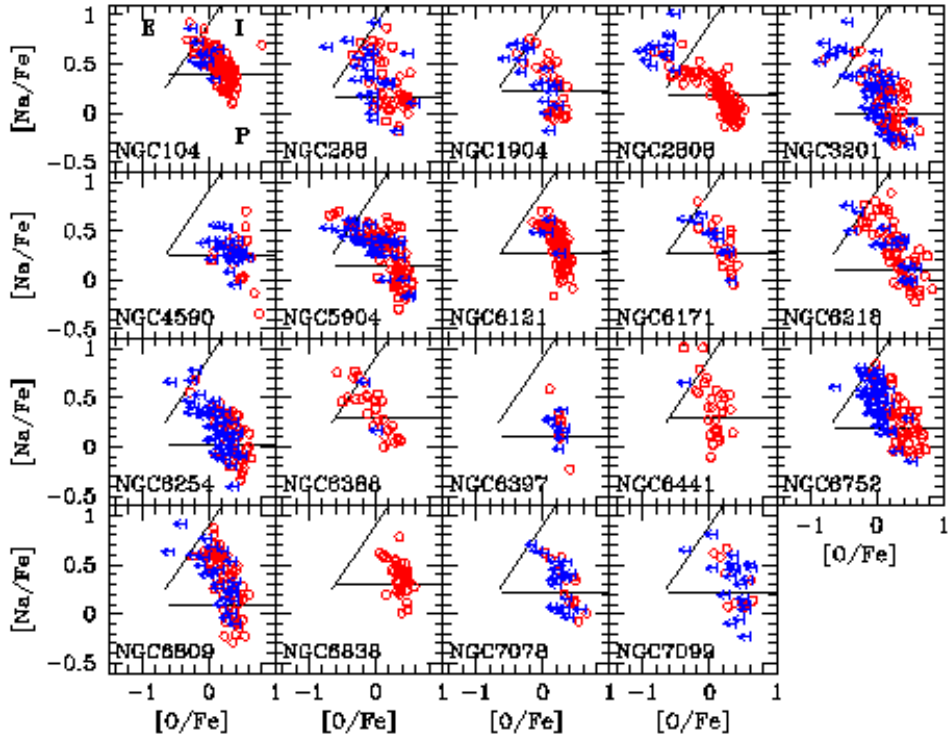} 
\caption{Na-O anticorrelations in 19 GCs observed with FLAMES@VLT, with
separation among Primordial, Intermediate, and Extreme populations (see
\cite[Carretta et al. 2009a]{C09a} for details).}
\label{figNaO}
\end{center}
\end{figure}

The work by Carretta and collaborators  (e.g. \cite[Carretta et al.
2009a]{carretta09a}) has further demonstrated the universality of the Na-O
anticorrelation; all GCs studied show it, as evident in the 19 GCs displayed in
Fig.\,\ref{figNaO}. However, this anticorrelation is not the same in all
clusters; the shape and the extension vary from cluster-to-cluster. On the basis
of the Na and O abundances, Carretta and collaborators separated the cluster
stars in first-generation and second-generation ones. The first are the ones
with Na and O similar to field stars of the same metallicity, the second show
varying degrees of O depletion and Na enhancement.   Na and O are not the only
light elements involved. Also Mg, Al (and even Si and F, \cite[Yong et al.
2005]{yong05}, \cite[Smith et al. 2005]{smith05}) are altered.  In particular,
Al is a powerful probe of the nature of first-generation polluters, due to the
very high temperatures required for its production.The   modification from the
primordial value varies a lot  from cluster to cluster (\cite[Carretta et al.
2009b]{carretta09b}: in some clusters Al changes a lot, in other much less or
not at all).  This is an indication that different polluters were at work. The
chemical changes come all from H-burning, but at very different temperatures,
and this indicates that polluters\footnote{While no definitive consensus has
been reached on the actual nature of the polluters, the most promising
candidates are fast rotating massive stars (e.g. \cite[Decressin et al.
2007]{frms}), and asymptotic giant branch stars (e.g., \cite[Ventura et al.
2001]{agb}). Since they are discussed in other contributions, I will not say
more on the subject. }  of different mass were at work in different GCs.

\section{The iceberg tip} A few clusters show the characteristics described
above in a extreme way and have been the footholds, if one may say so, to
convince even the distracted astronomer that something did not fit the notion of
GCs being simple, old, boring systems.

\subsection{$\omega$ Centauri} 
\begin{figure}[]
\begin{center}
\includegraphics[bb=170 330 430 520,clip, width=4cm]{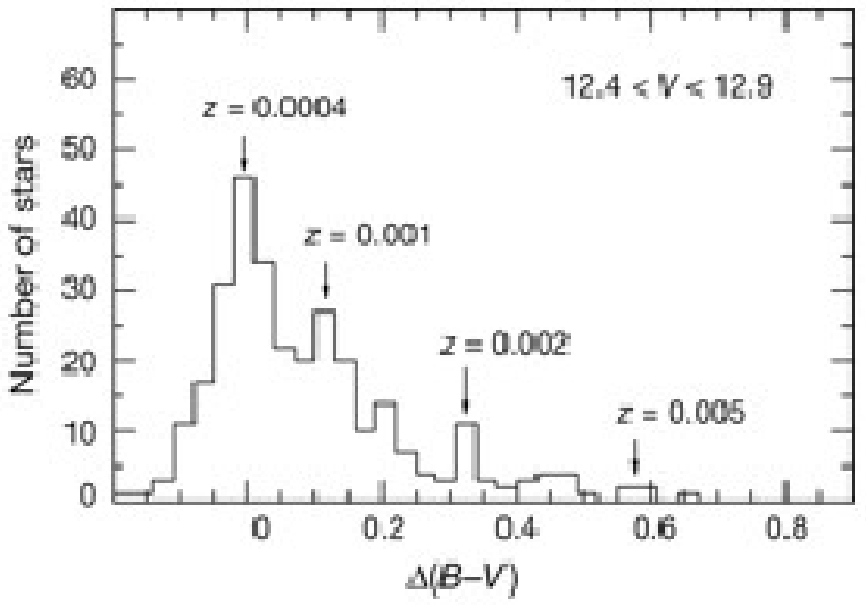} 
\includegraphics[width=4cm]{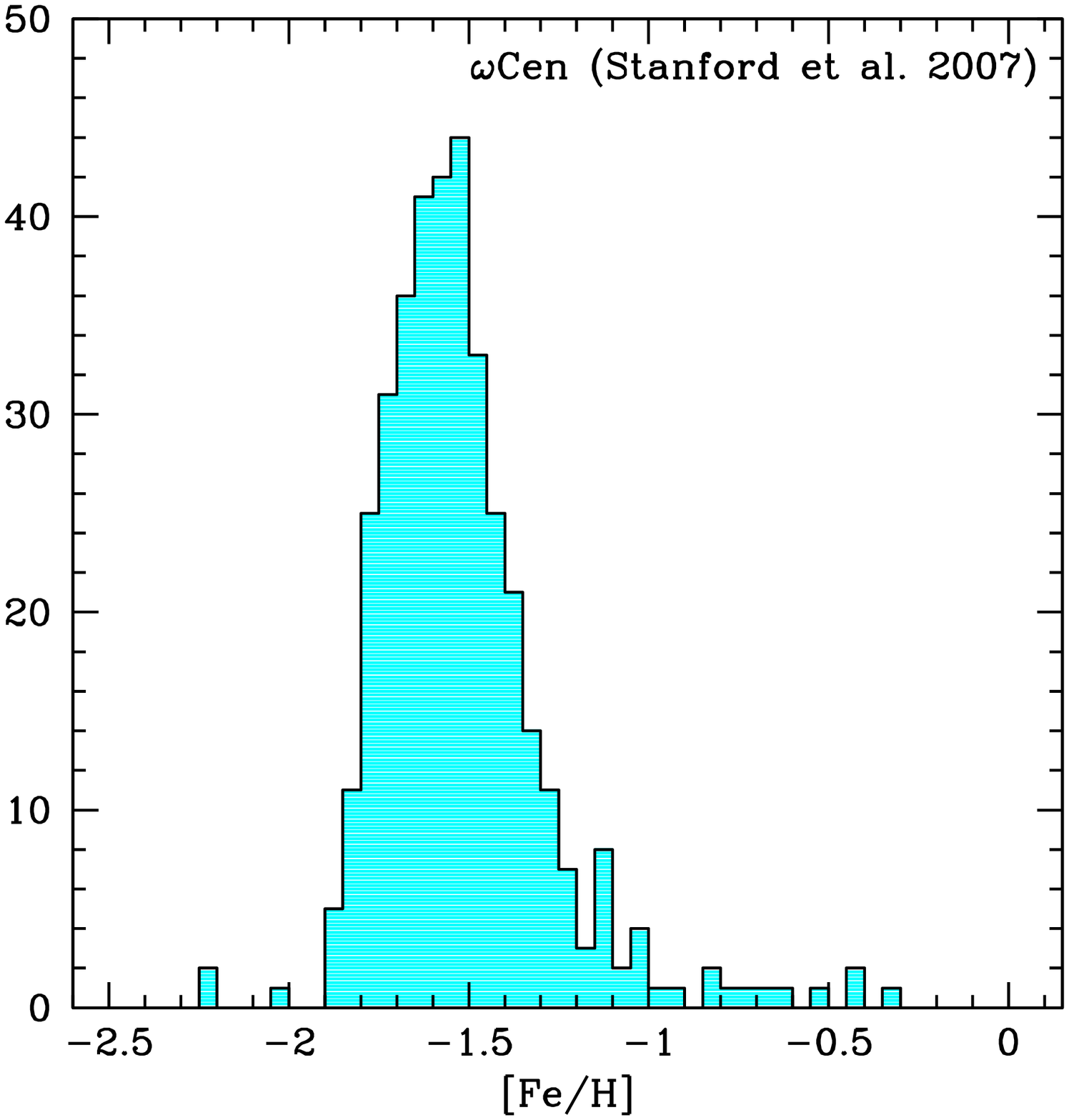}
\includegraphics[width=4cm]{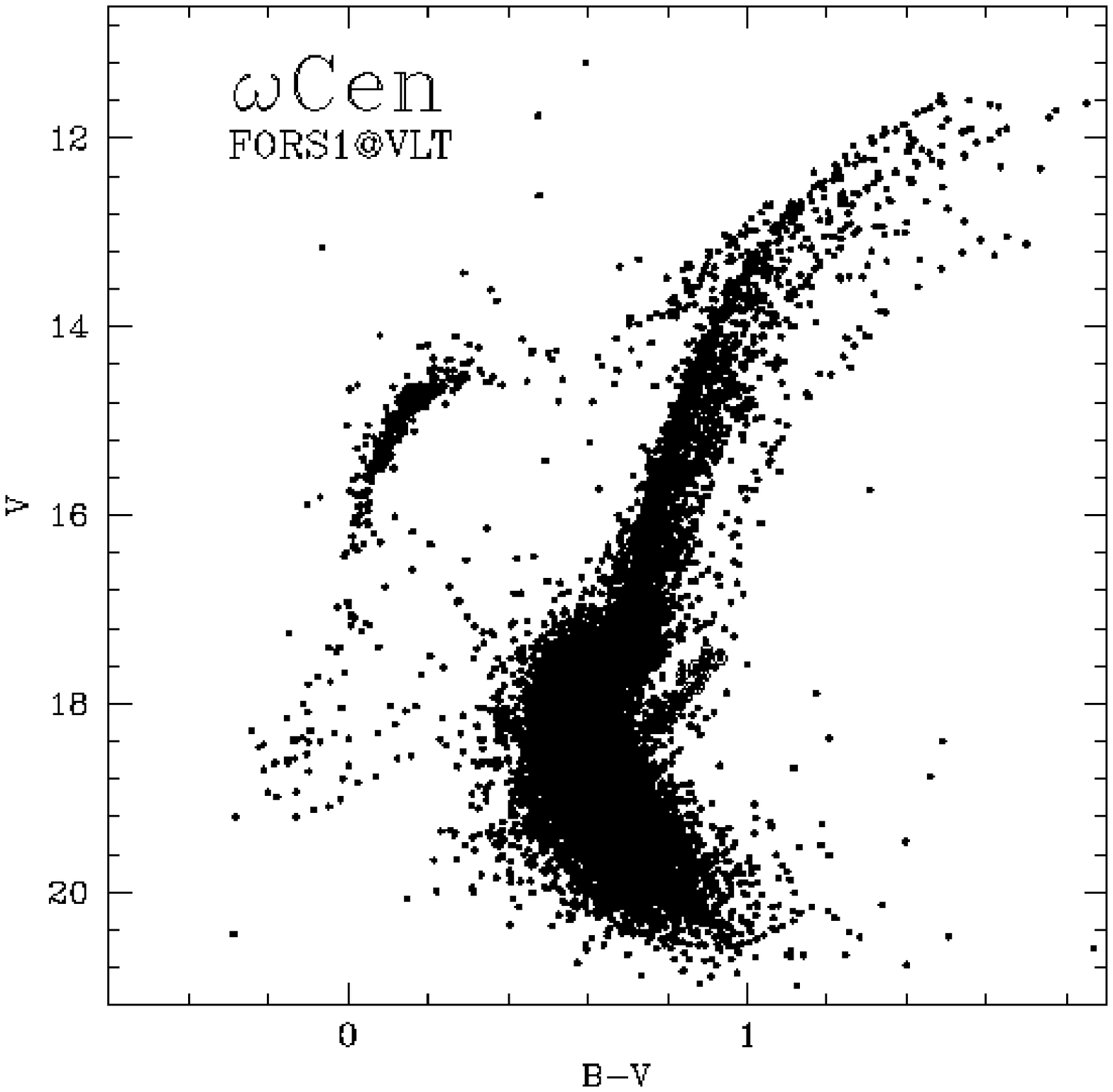}
\includegraphics[bb=20 270 570 600,clip,width=10cm]{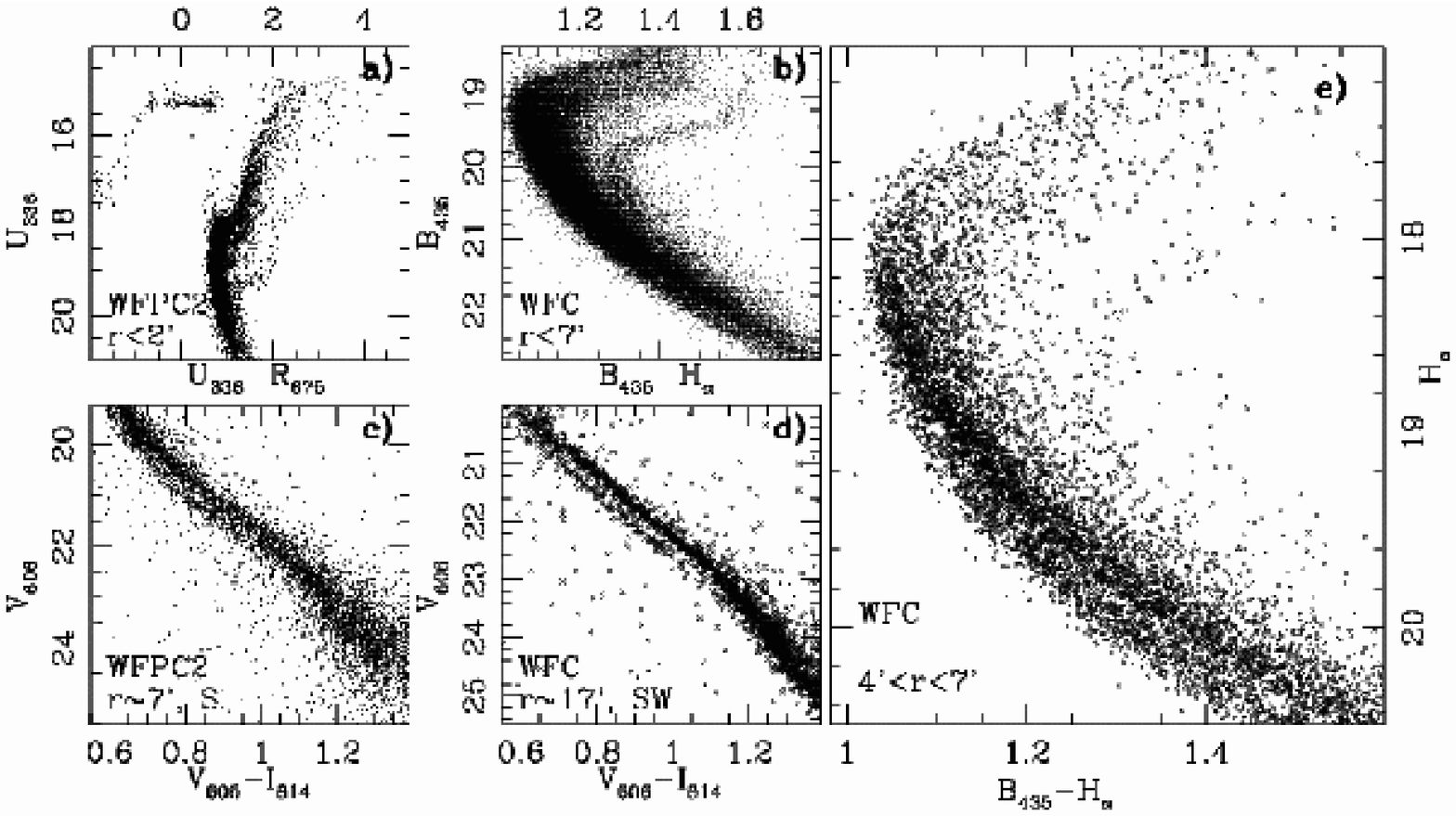}
\caption{Evidence of multiple populations in $\omega$ Cen: Upper left panel:
distribution in colour of RGB stars from \cite[Lee et al. (1999)]{lee99}.
Central upper panel: distribution in [Fe/H] of main sequence stars, adapted
from \cite[Stanford et al. (2007)]{stanford07}. Upper right panel: 
distinct RGBs, from \cite[Ferraro et al. (2004)]{ferraro04}. Lower panel:
collection of CMDs showing the various populations, both for evolved and main
sequence stars, from \cite [Bedin et al. (2004)]{bedin04}.}
\label{figomega}
\end{center}
\end{figure}

When talking of the complexity of GCs, the first
example that comes to mind is of course $\omega$ Cen.   It had long been
suspected that is could host stars of different metallicity, given the width of
the sequences (e.g.,  \cite[Cannon \& Stobie 1973]{cs73},  \cite[Alcaino \&
Liller 1987]{al87}) and also demonstrated by pioneering spectroscopic work  (see
\cite[Butler et al. 1978]{but}, \cite[Cohen 1981]{cohen81}). Given the huge
amount of papers dedicated to $\omega$ Cen, I can only touch upon a tiny
fraction of the works on this cluster, concentrating on the very recent results,
which show not only dispersion in colour or metallicity, but actual discrete
separation in different sub-samples of the total cluster population. Initially
the clear indication of multiple populations came from the RGB (\cite[Lee et al.
1999]{lee99}, \cite[Pancino et al. 2000]{pancino00}, \cite[Ferraro et al.
2004]{ferrarp04}), see Fig.\,\ref{figomega}. Several metallicities, and perhaps
different ages were deduced for the different RGB sequences, in particular for
the very metal-rich ``anomalous RGB", see for instance the high resolution
spectroscopic study of 40 giants by \cite[Norris \& Da Costa (1995)]{ndc95}, the
work on Ca abundances of about 500 giants observed at low spectroscopic
resolution by \cite[Norris et al (1996)]{norris96},  the Str\"omgren photometry
by \cite[Hilker \& Richtler (2000)]{hr00}, the high resolution analysis of 6
stars on the ``anomalous" RGB by \cite[Pancino et al. (2002)]{pancino02}, the
near-IR spectroscopy of about 20 giants by \cite[Origlia et al.
(2003)]{origlia03}, the  study of main sequence stars by \cite[Stanford et al.
(2007)]{stanford07}, the large survey at intermediate spectroscopic resolution
by \cite[Johnson et al. (2009)]{johnson09}; see also Marino et al. at the
present conference.

However, the differences go deeper, and with data obtained with HST,  it was
possible to detect also separate main sequences (\cite[Bedin et al.
2004]{bedin04}, see Fig.\,\ref{figomega}), to which I will come back later.

\subsection{NGC 2808}
\begin{figure}[]
\begin{center}
\includegraphics[bb=10 140 550 680, clip, width=5cm]{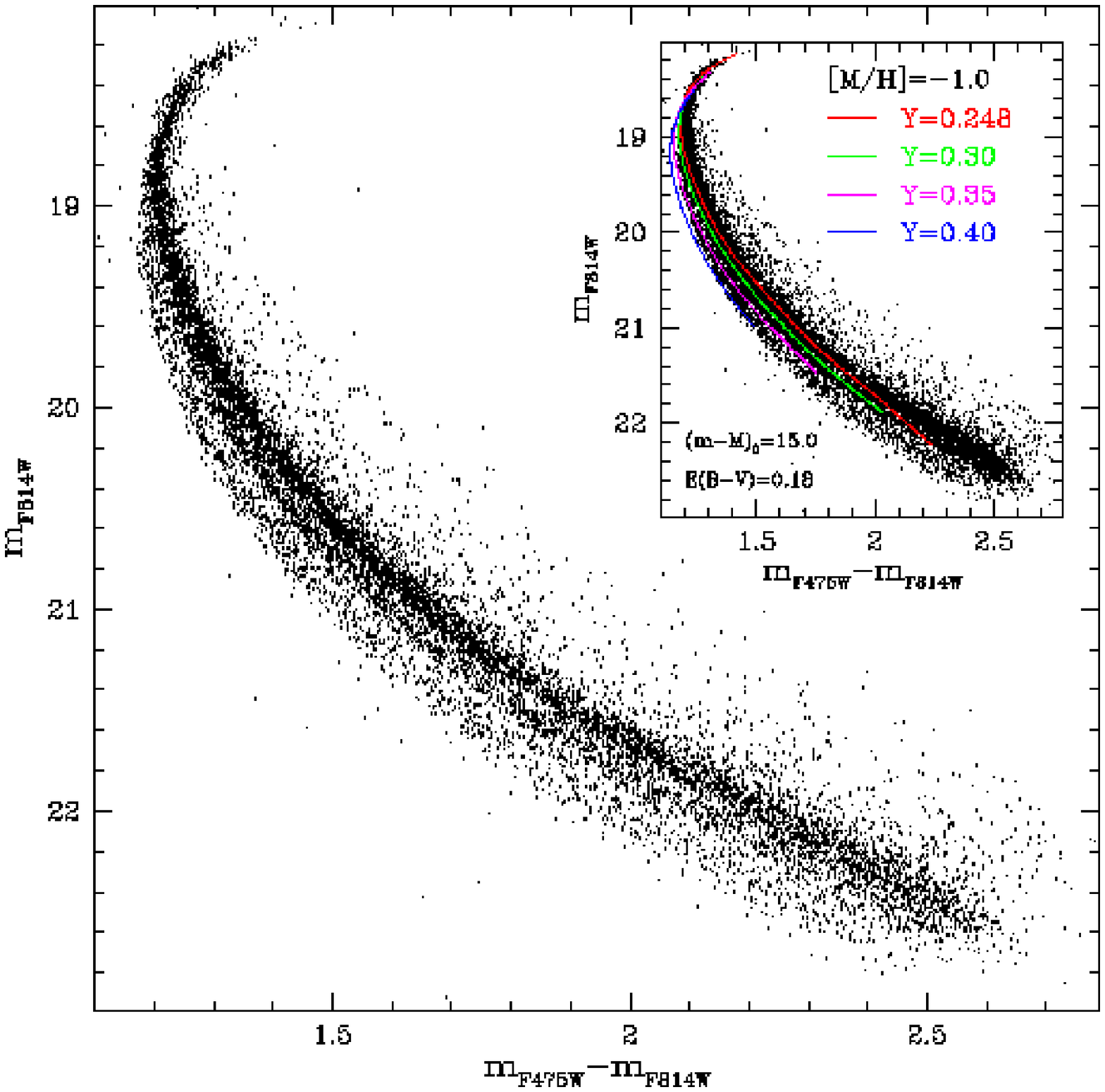} 
\includegraphics[bb=10 120 570 690, clip,width=4cm]{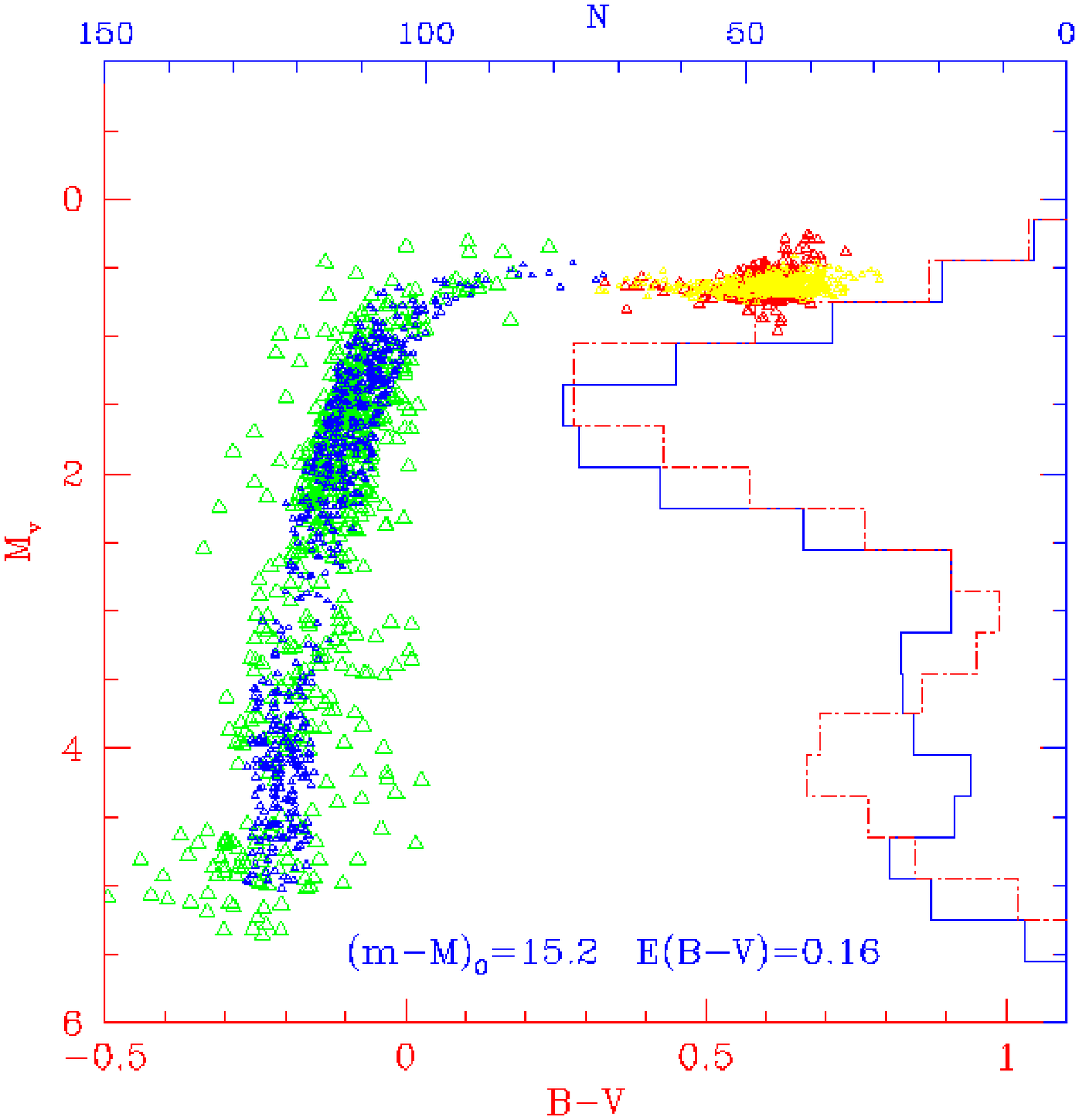} 
\includegraphics[bb=10 140 550 680, clip,width=4.25cm]{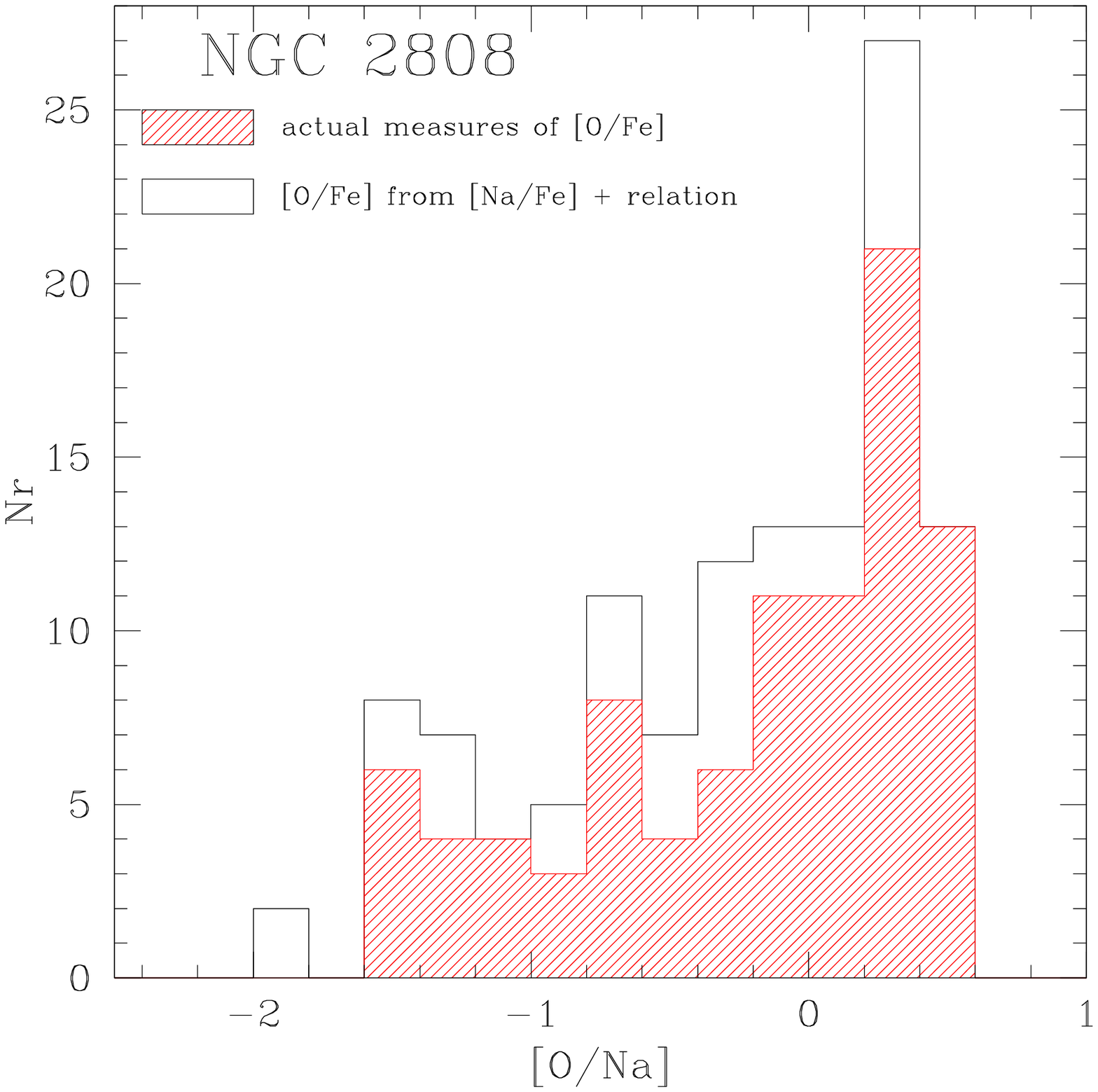} 
 \caption{Evidence of multiple populations in NGC 2808. Left panel: the three
separate main sequences found by \cite[Piotto et al. (2007)]{p8n2808}. Central
panel: the complex HB, and the interpretation assuming three different He
contents, made by \cite[D'Antona et al. (2005)]{dantona05}. Right panel: the
distribution of Na and O abundances, with three peaks, as found in 
\cite[Carretta et al. (2006)]{carretta06}.}
\label{fig2808}
\end{center}
\end{figure}

Even more ``normal", less massive clusters show striking features. NGC 2808,
among many other marked peculiarities, presents three well separated main
sequences (\cite[Piotto et al. 2005]{p805}). It also has a very complex
Horizontal Branch (HB), with three main groups of stars, that cannot be
explained under standard assumptions. Both these features seem to require He
enhancement in part of its stars (see next Section), which would also  naturally
agree very well with the observed Na-O anticorrelation (the third most extended
one, after $\omega$ Cen and M54), since  Na-rich and O-poor stars should also be
He-rich (the main outcome of H-burning is, of course, He).  Fig.\,\ref{fig2808}
shows the three main sequences, a plausible interpretation of its HB
(\cite[D'Antona et al. 2005]{dantona05}), and the Na-O anticorrelation
(\cite[Carretta et al. 2006]{carretta06}).

\subsection{M54}
\begin{figure}[]
\begin{center}
\includegraphics[bb=170 310 410 550, clip, width=4cm]{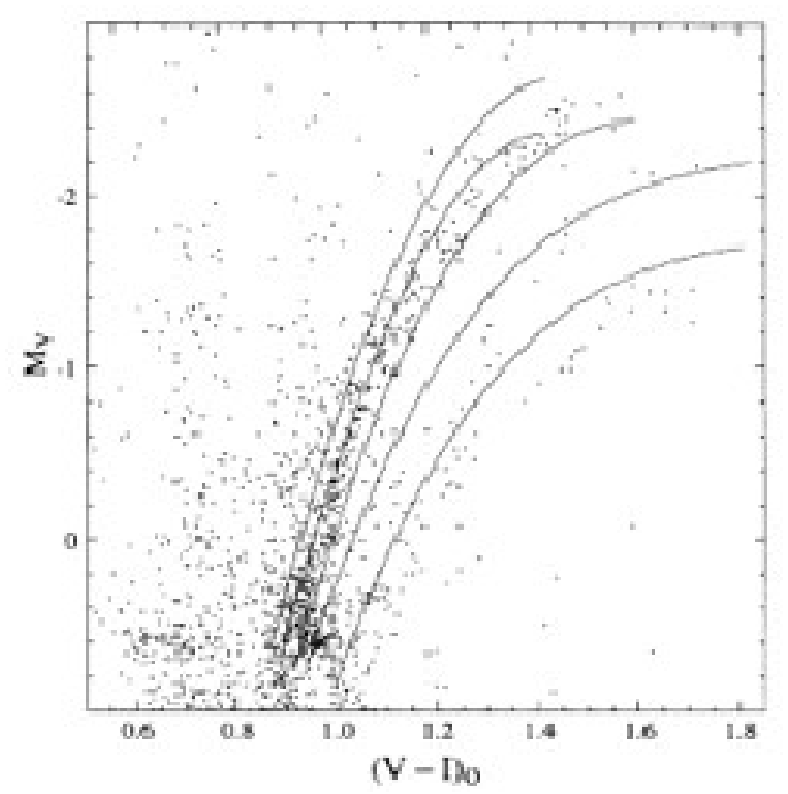} 
\includegraphics[bb=30 40 412 570, clip, width=5cm]{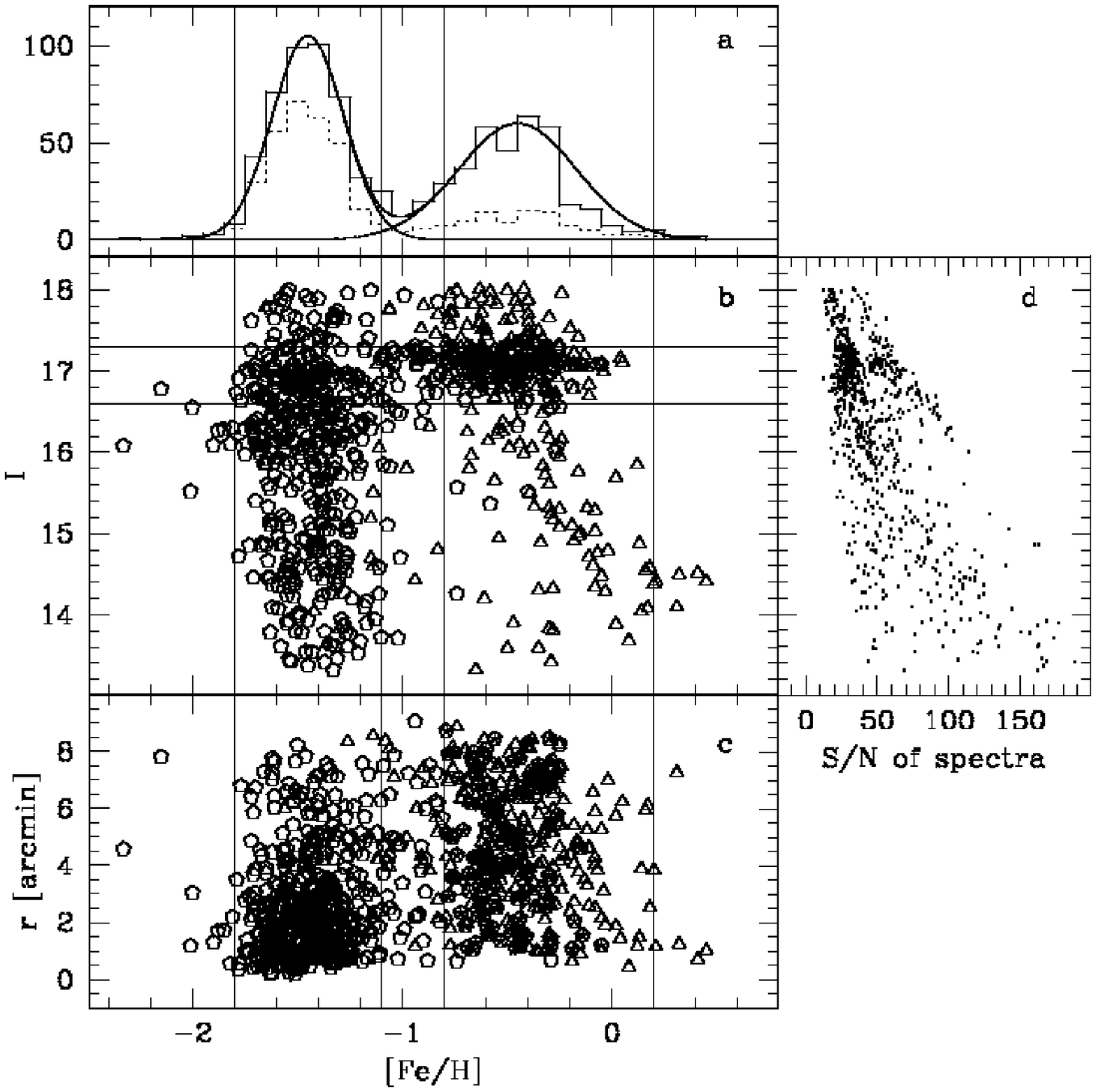} 
\includegraphics[bb=20 430 300 710, clip,width=4cm]{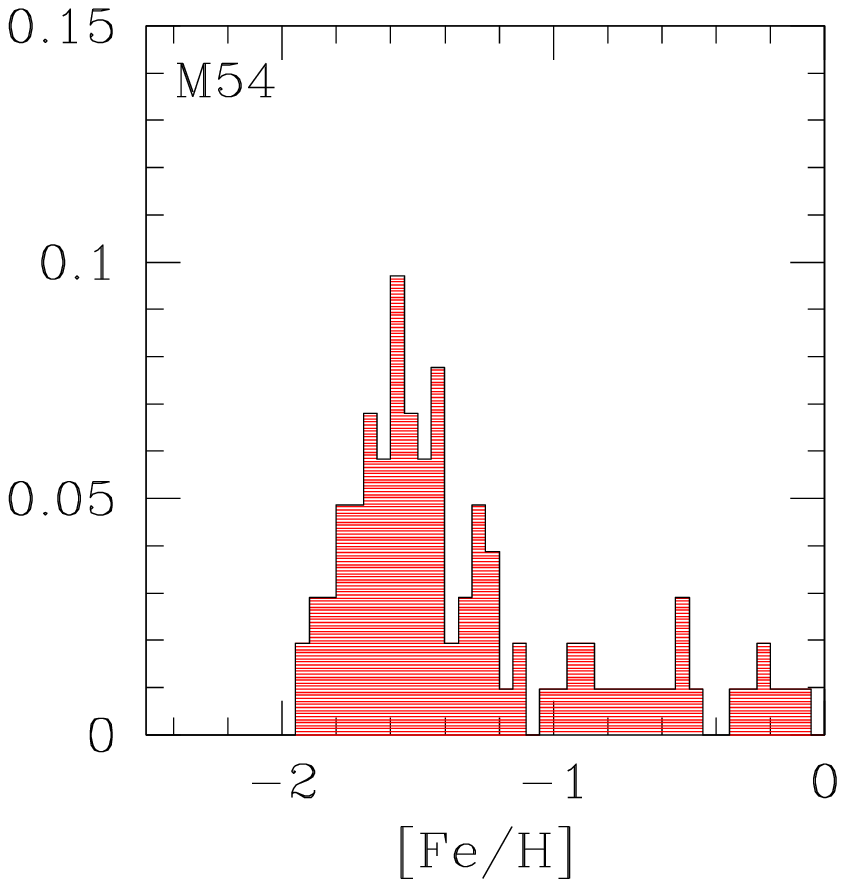} 
\caption{Dispersion in metallicity in M54. Left panel: \cite[Sarajedini \&
Layden (1995)]{sl95} found a colour dispersion in the RGB. Central panel:
Separation of M54 from the Sgr population in  \cite[Bellazzini et al.
(2008a)]{bella08a}, based on Calcium triplet observations of several hundreds of
stars; notice that both M54 and Sgr show a noticeable dispersion in [Fe/H].
Right panel: results obtained by \cite[Carretta et al. (2010)]{carretta10} from
76 star in M54 (peaked at [Fe/H]$\sim-1.6$) and 25 stars in Sgr (extending in
[Fe/H] from about $-1$ to solar) observed with FLAMES@VLT.}
\label{figm54}
\end{center}
\end{figure}

M54 is the second most  massive cluster of the  Galaxy, and lies at the center
of the disrupting Sgr dwarf galaxy. It has been suspected to have a dispersion
in metallicity since the observations by \cite[Sarajedini  \& Layden
(1995)]{sl95}, whose CMD shows a wide RGB, compatible with a dispersion of 0.16
dex in [Fe/H], see Fig.\,\ref{figm54}. This has been recently confirmed by low
resolution spectroscopy of a very large sample of M54 and Sgr stars
(\cite[Bellazzini et al. 2008a]{bella08a}, see Fig.\,\ref{figm54}). The very
recent results obtained by \cite[Carretta et al. (2010)]{carretta10} using
FLAMES spectra of about 80 RGB stars further confirm this: M54 has a dispersion
in metallicity of the order of about 0.2 dex, well above the errors (see
Fig.\,\ref{figm54}). Furthermore, it has a very extendend Na-O anticorrelation,
more extended for the metal-rich than for the metal-poor stars. \cite[Carretta
et al. (2010)]{carretta10} also noticed that the same happens in $\omega$ Cen.
M54 deserves more study, but it's clear that it resembles $\omega$ Cen; maybe,
as it has been suggested (\cite[Bellazzini et al. 2008a]{bella08}, 
\cite[Carretta et al. 2010]{carretta10}), we see it now as $\omega$ Cen was a
long time ago, before the dwarf galaxy around it dispersed.

\section{Are there different He abundances in GCs ?}

The main problem for establishing if He is  variable in GCs it that it is
difficult to see the effect of small variations in the CMDs, apart from  the HB,
which is a sort of ``amplifier".  HBs suffer from the notorious
``second-parameter" problem: metallicity is of course the first parameter
explaining their structure, age is most probably the second, but their
combination cannot explain all HBs, in particular at the bluer, hotter
extremes.  

A solution is to bring also He into the problem, since an higher He content
means brighter and bluer HBs.  As shown  in Fig.\,\ref{fig2808}, \cite[D'Antona
et al. (2005)]{dantona05} were able to reproduce the observed distribution of HB
stars in NGC 2808 assuming three different He contents (from a ``primordial"
value of Y=0.25 for the red HB, to Y=0.40 for the extreme blue HB). A similar
exercise has been done by \cite[Busso et al. (2007)]{busso} and \cite[D'Antona
\& Caloi (2008)]{dc08} for  NGC 6388, producing similar results. This is a
strong indication that GC stars are not so homogeneous in He as we thought: in
the same cluster we may have stars with ``normal" He and stars with different
levels of He-enhancement, even if  perhaps this doesn't happen in all clusters,
or at least not at this  high level. 

It could be very interesting to directly measure He abundances in GC stars;
however, this is possible only on the HB, and with many limitations. There is a
small range in temperature (9500-11500 K) where He lines can be observed (even
if they are tiny, at a few percent of the continuum level) and He abundances are
not altered by dilution and mixing.  Recently, \cite[Villanova et al.
(2009)]{villanova} have obtained high resolution, high S/N of  a few HB stars in
NGC 6752, but their results are not decisive: they could measure He only in four
stars, all of them Na-poor, O-rich (hence expected to be He-``normal", that is
what they found), while they could not do so for the only Na-rich, O-poor star
(expected to be He-enhanced). On the other hand,  He has been found to be
enhanced in some very blue HB stars in NGC 2808 and $\omega$ Cen (see
\cite[Moehler et al. 2007]{mo07} and Fig.\,\ref{figomega}), even if the cause of
He-enhancement could also be mixing following a late He-flash (\cite[Castellani
\& Castellani 2003]{cc}) for these extreme HB stars.

In the present volume there is also a discussion (see \cite[Bragaglia et al.
2010]{B10} for a lengthier presentation) on how to determine  differences in He
abundances from RGB stars using colours, metallicities, and magnitude of the RGB
bump. 

Finally, maybe the strongest case for He-enhancement of part of GC stars comes
from two massive objects. The first one is again $\omega$ Cen, with its two
separate main sequences (see Fig.\,\ref{figomega}). \cite[Piotto et al.
(2005)]{p8omega} obtained spectra of moderate resolution for 17 stars on the
blue and 17 stars on the red sequences; surprisingly, the blue sequence turned
out to be more metal-rich by about 0.3 dex. This could be explained only if we
assume that the blue sequence is also much more He-rich than the red one
(Y$\sim$0.35-0.38 $vs$ 0.25) as seen from  isochrone fit. Of course we have to
remember that the different metallicities are actually measured, while the
He-enhancement is only inferred.

The split main sequence is even more spectacular in NGC 2808
(Fig.\,\ref{fig2808}), and again the three sequences can be well fit assuming
the same age but three different He levels (i.e., Y$\sim$ 0.25, 0.30, and 0.38,
see \cite[Piotto et al. 2007]{p8n2808}).
 
\begin{figure}[]
\begin{center}
\includegraphics[bb=20 160 550 680, clip,width=5.5cm]{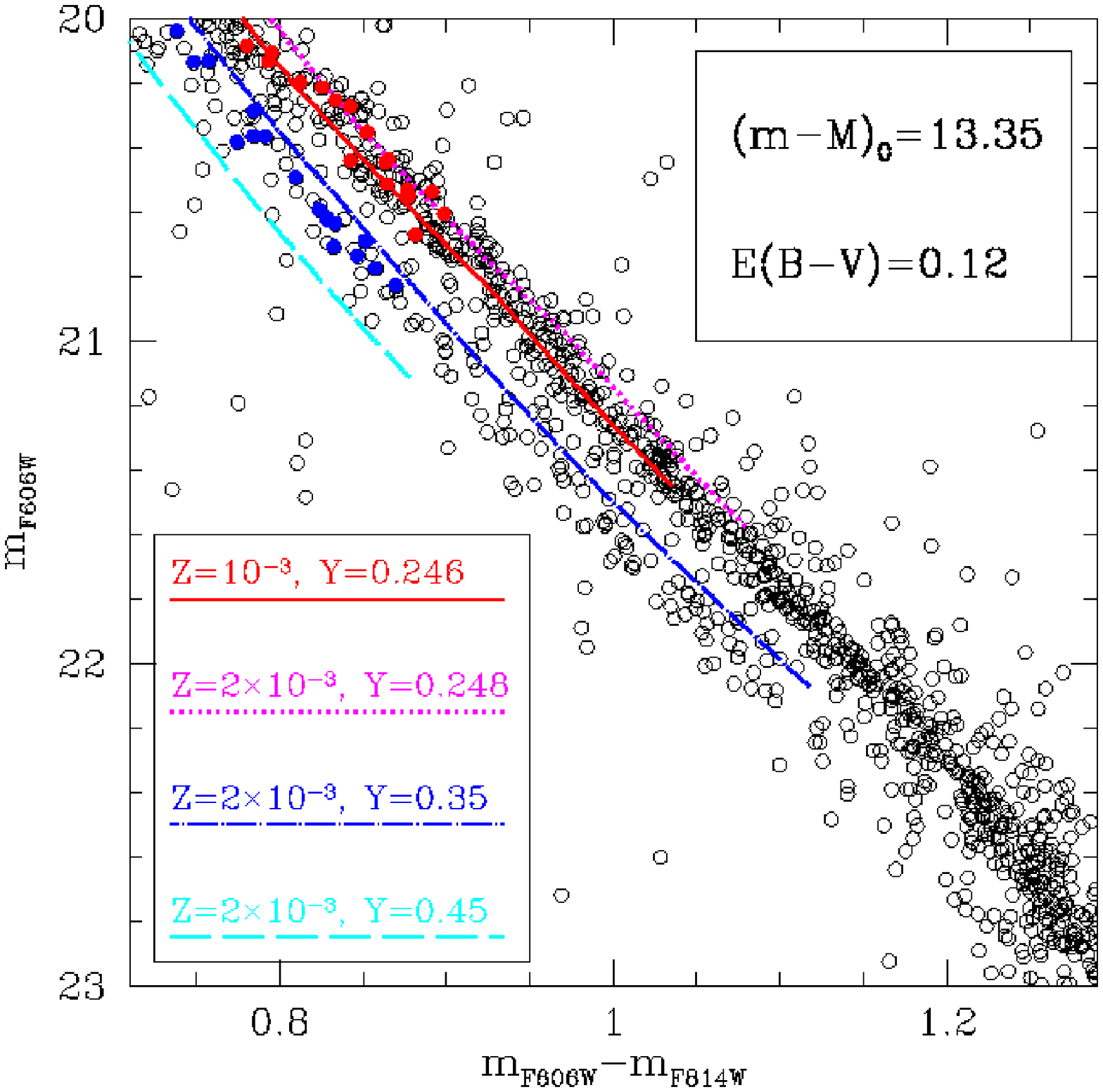} 
\includegraphics[bb=30 230 530 610, clip,width=5.5cm]{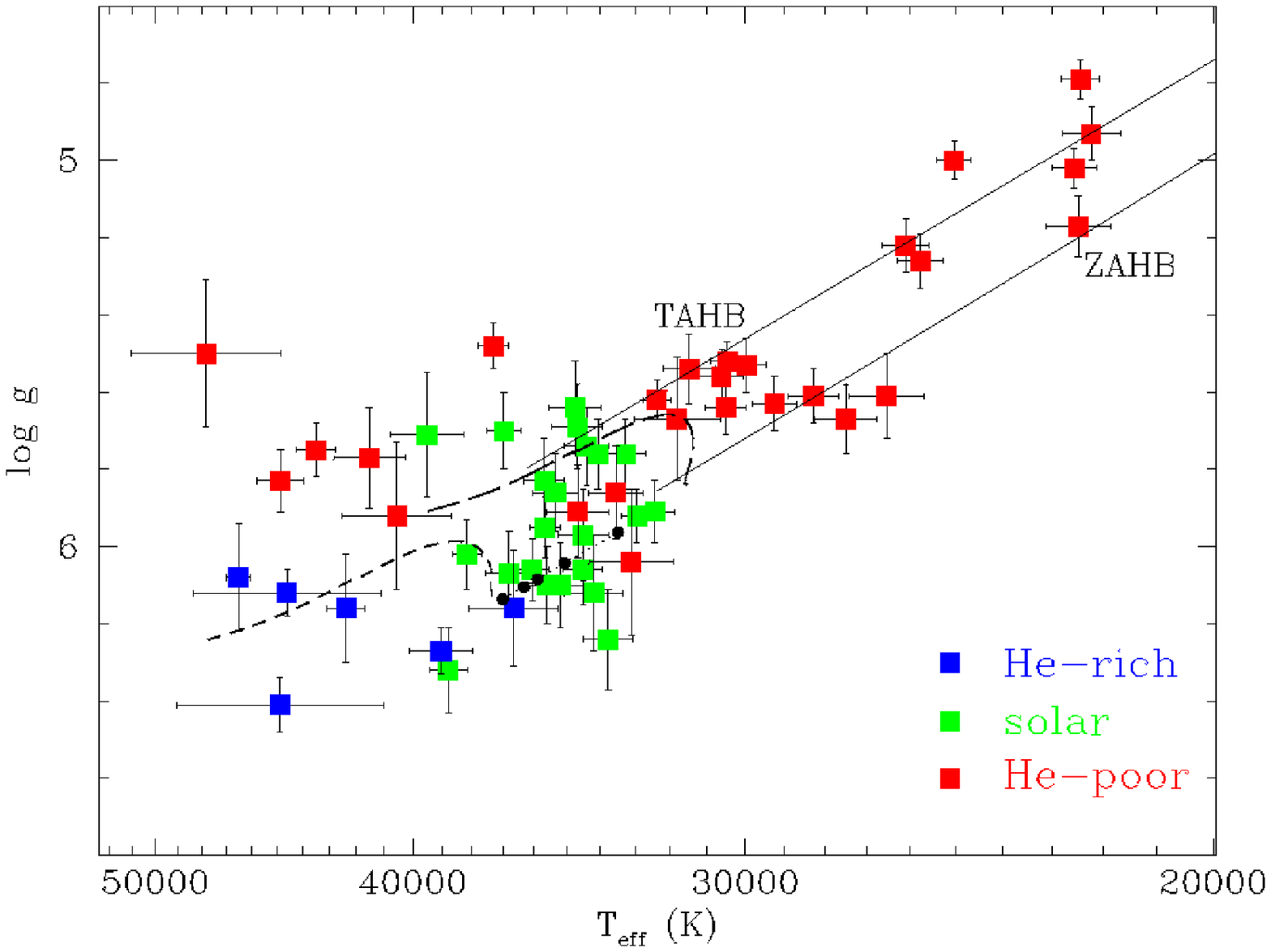} 
\caption{He abundances in $\omega$ Cen. Left panel: the two main sequences in
\cite[Piotto et al. (2005)]{p8omega}, with star observed with FLAMES indicated
by filled red and blue circles. The isochrones shown have been computed for
Z=$1\times10^{-3}$, Y=0.25 and Z=$2\times10^{-3}$, Y=0.35. Right panel: blue and
extreme HB stars for which \cite[Moehler et al. (2007)]{mp07} obtained low
resolution spectra and for which determined temperatures, gravities, and He
abundances; part of the extreme HB stars are He-enhanced.}
\label{figHe}
\end{center}
\end{figure}

\begin{figure}[]
\begin{center}
\includegraphics[width=8cm]{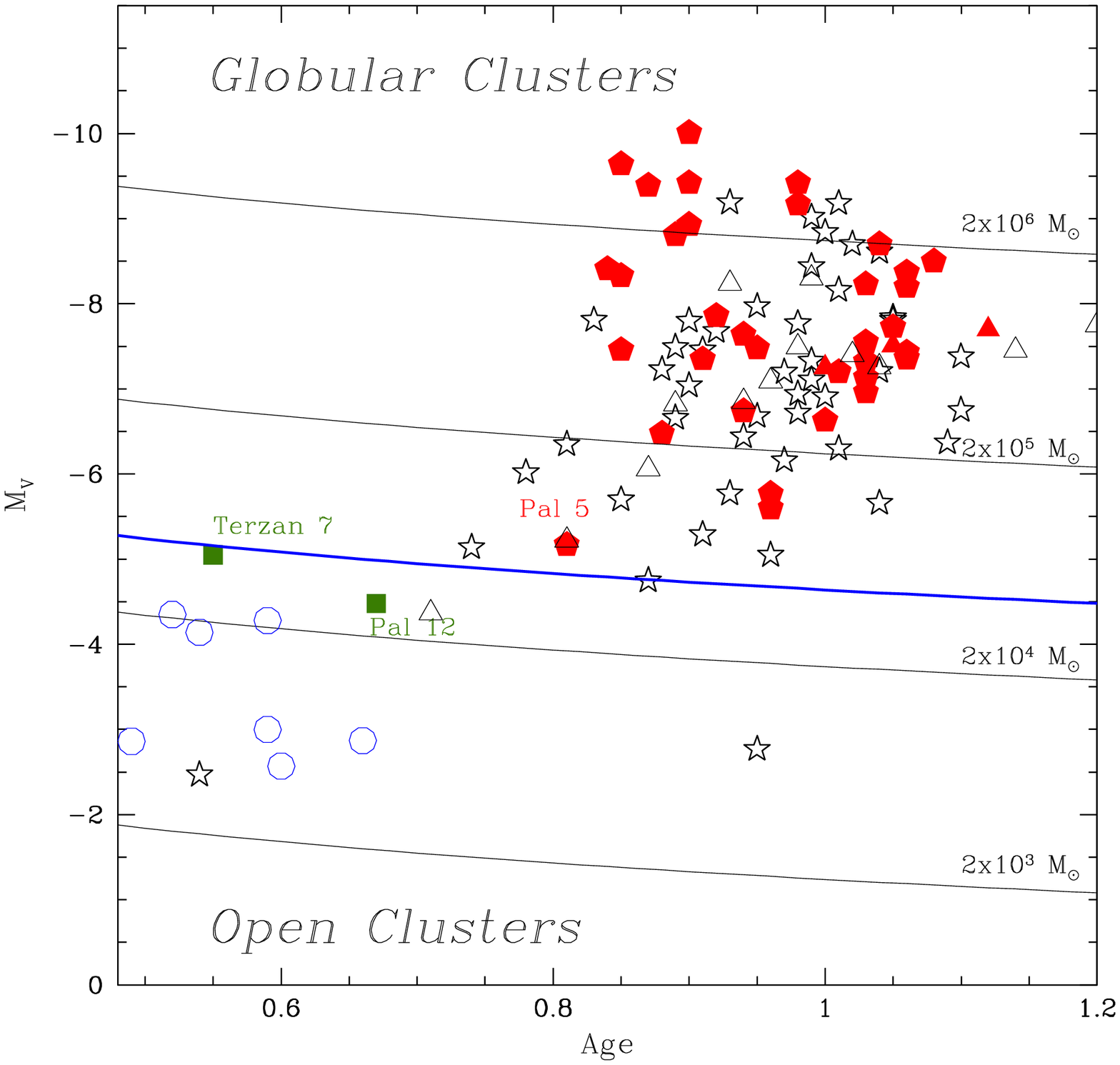} 
\caption{Relative Age parameter vs absolute magnitude $M_V$ for globular and 
old open clusters. Red filled pentagons and triangles  are GCs where Na-O
anticorrelation has been observed, in the Milky Way or the  LMC respectively;
green squares are clusters which do not show (yet?) evidence  for O-Na
anticorrelation (Terzan 7 and Pal 12,  both in the Sgr dwarf  galaxy). Open
stars and triangles mark clusters for which not enough data is available.
Finally, open circles are old open  clusters (data from \cite[Lata et al.
2002]{lata}). Superimposed are lines of constant mass (light solid lines, see
\cite[Bellazzini et al. 2008b]{bella08b}). The heavy blue solid line  (at a mass
of $4\times 10^4~M_\odot$) is the proposed separation between globular and open
clusters. This figure is taken from Carretta et al., submitted }
\label{figzcorri}
\end{center}
\end{figure}

\section{Summary and perspectives}
We have seen that there is photometric evidence of multiple populations in many
GCs. This  generally happens among the most massive ones in our Galaxy, but not
exclusively (see the cases of NGC 6752, M4). Mass is an important factor. It has
been shown that many properties correlate with cluster mass, for instance, the
maximum temperature reached on the HB  (\cite[Recio-Blanco et al. 2006]{ale})

However, mass is not all the story. We have seen that all GCs display the Na-O
anticorrelation, but if we quantify its extension (using e.g., the
Interquartile  range, see \cite[Carretta 2006]{eu}) and plot it against total
cluster mass, or integrated magnitude, we see that, preferentially,  only
high-mass clusters have an extended anticorrelation. This is, however,  only a
necessary condition; a notable counterexample is the  massive GC  47 Tuc, which
has a short  anticorrelation. Some other factor, maybe metallicity, or age, or
cluster orbit have to be involved.

Finally, I recall that the Na-O (and similar) anticorrelations seem to represent
an intrinsic property of GCs: each time Na and O have been measured,  they
anti-correlate, while this does not happen in open clusters (see
Fig.\,\ref{figzcorri}) or for field stars. So maybe we have an operative
definition of the separation between globular and open clusters: GCs are those
aggregates massive enough to sustain self-pollution, hence able to host at least
two stellar generations and to develop  a Na-O anticorrelation. This has of
course to be related to the mechanism of cluster formation.

\bigskip The financial support from the IAU and  the Swiss National Science
Foundation is gratefully acknowledged. Many thanks to Eugenio Carretta, Raffaele
Gratton, Valentina D'Orazi, Sara Lucatello, and Antonino Milone for their help
and suggestions.

\end{document}